\documentclass[letterpaper, 10 pt, conference]{ieeeconf}  

\IEEEoverridecommandlockouts                              
\usepackage{defs}

\newcommand{\non}{\nonumber}
\newcommand{\mbb}[1]{\mathbb{#1}}
\newcommand{\mcal}[1]{\mathcal{#1}}
\newcommand{\W}{\widehat{\mbb{W}}}

\title{\LARGE \bf
An Adaptation of the AAA-Interpolation Algorithm for Model  Reduction  of MIMO Systems
}

\author{Jared Jonas and Bassam Bamieh%
\thanks{Jared Jonas and Bassam Bamieh are with the Department of Mechanical Engineering, University of California - Santa Barbara, USA 
{\tt\small jjonas@ucsb.edu, bamieh@ucsb.edu}}%
}

\begin{document}
\maketitle
\thispagestyle{empty}
\pagestyle{empty}

\begin{abstract}
We consider the Adaptive Antoulas-Anderson (AAA) rational interpolation algorithm recently developed by Trefethen and co-authors,  
which can be viewed as a type of moment-matching technique for system realization and approximation. We consider variations 
on this algorithm that are suitable for model reduction of linear time invariant systems while addressing some of the shortcomings 
of the block-AAA variant of the algorithm for MIMO systems. In particular, we develop state-space formulas and keep track of the state-space dimension at every step of the adaptive block-AAA algorithm, showing an unfavorable increase of the state dimension.  We propose  a new low-rank adaptive  interpolation algorithm that addresses this shortcoming. 
Comparative  computational results are included for the algorithms above, together with comparisons to balanced reduction.  
\end{abstract}

\section{Introduction}
Model order reduction is an important tool in the analysis, simulation, and control of large-scale systems~\cite{antoulas2000survey,Baur14}, and is particularly relevant for control applications in, for example, fluid and structural mechanics~\cite{Lassila14,Hetmaniuk12}.  In the context of linear dynamic systems, model reduction algorithms aim to produce a state-space model with fewer states that approximates the dynamics of the original system.  
Amongst several model-reduction techniques, moment matching  constructs a reduced-order model that matches the original model's moments at a given set of points~\cite{ionescu_astolfi_2011}.  This can be interpreted  as creating a rational interpolant whose value (or some derivative) matches the original transfer function at that point. Moment matching and interpolation problems are therefore intimately linked. 

Building on the original rational interpolation results of Antoulas and Anderson~\cite{antoulas1986scalar} 
that uses a 
pre-specified set of interpolation points, Trefethen et. al~\cite{Nakatsukasa_2018} developed an algorithm they termed Adaptive Antoulas-Anderson (AAA). This algorithm uses  a barycentric interpolation formula~\cite{berrut_trefethen_2004}, and ``adaptively'' picks points in the complex plane at which a scalar-valued  function is interpolated based on a maximum error criterion. The algorithm yields a rational approximant to a given complex function. 
Its main advantage is the automated selection of interpolation points, and has several interesting features as discussed in~\cite{Nakatsukasa_2018}. 

Subsequently, a matrix-valued version of the algorithm, termed  
 block-AAA~\cite{gosea2021algorithms} was developed. This algorithm interpolates the {\em matrix value} of a given function at certain points that are also  adaptively selected according to a maximum error criterion. 

Since their introduction, AAA and related algorithms have been used in a systems context for  model-order reduction and also in system identification.  
Such ``data-driven'' rational approximations have been used in  parametric dynamical systems~\cite{Rodriguez23}, and in quadratic-output systems~\cite{gosea2021datadriven}.  More recently, they 
have been used in a model-order reduction scheme~\cite{yu2023leveraging} with a two-step method  utilizing both block-AAA on a discrete set of points and Hankel norm approximation.

In this paper, we propose new variants of the AAA algorithm for the purpose of model reduction of high-order LTI systems. We give state-space formulas for realizations of interpolants with real parameters. We also replace the discretized maximum criterion employed in previous algorithms by a bisection algorithm for computing $L^\infty$ errors on the imaginary axis, which in turn guides the adaptive selection of interpolation points. Most importantly, we show that adapting the existing block-AAA algorithm for use on linear systems has undesirable features when used on MIMO systems, especially when the number of outputs is large, in that it leads to a rapid increase in the state dimension of the interpolant compared to other schemes. The requirement of exactly interpolating the full matrix at each point causes this increase in state dimension. We argue that matrix-valued interpolation with lower rank matrices (formed from the significant singular values/vectors at those points) rather than exact interpolation is more effective. With this motivation, we develop an algorithm and demonstrate its effectiveness with numerical examples comparing the proposed algorithms with balanced reduction. We close with a discussion of some open problems in matrix-valued interpolation, and directions for future work. 


\subsection{Notation}
We use the notation 
\[
	H(s) = C(sI-A)^{-1} B + D =  \brac{\begin{array}{c|c}A & B \\ \hline C & D\end{array}}
\]
for the transfer function and state space realization of a finite-dimensional Linear Time Invariant (LTI) system. 
\(\overline{X}\) denotes the complex conjugate (not transpose) of  a matrix \(X\), and  \(X^*\) denotes the complex-conjugate transpose of \(X\).  

\section{System-AAA}
The block-AAA algorithm~\cite{gosea2021algorithms} is an iterative algorithm that starts with a given matrix-valued function $G(.)$ (of possibly high order), 
and builds up a matrix-valued rational function approximation at step $r$  of the form
\begin{align}
        R_r(z) &= \p{\sum_{k=1}^r \frac{W_k}{z-z_k}}\inv \p{\sum_{k=1}^r \frac{W_kG(z_k)}{z-z_k}} 	  \label{blockAAA}	\\
        &=: M^{-1}_r(z) ~N_r(z) . 															\nonumber
\end{align}
This particular form ensures that $R_r$ interpolates $G$ exactly at the so-called support points $\left\{ z_k \right\}$ in the sense
that $R_r(z_k) = G(z_k)$ as matrices. The weight  matrices $\left\{ W_k \right\}$ are free parameters 
 chosen to minimize some measure (usually a least squares criterion) of 
error between $R_r$ and $G$ over (typically a large number of) points in the domain $\Omega$. 
The  next support point $z_{r+1} \in \Omega \subset \mathbb{C}$ is 
chosen where  the following error criterion is maximized 
\begin{equation}
	z_{r+1} = \arg\min_{z\in\Omega} \left\| R_r(z) - G(z) \right\| .
  \label{next_support}
\end{equation} 
The rationale being that since interpolation is exact at the support points, this error will be most reduced by this choice at 
the next iteration. 

The block-AAA algorithm presented in~\cite{gosea2021algorithms} produces approximations that have complex coefficients, and only evaluates the least squares error and solves the problem~(\ref{next_support}) numerically  over a large grid of points in $\Omega$. In this section, we propose a variant we call system-AAA, which works directly with state-space realizations with real matrices, performs the support point selection step~(\ref{next_support}) using a bisection algorithm (similar to those for computing $H^\infty$ norms), and selects the matrix weights $\left\{ W_k \right\}$ using a solution of the least squares problem without gridding. The solution of this last problem involves computing Gramians of systems and finding eigenvectors of matrices related to them. Thus gridding of the domain $\Omega$ is completely avoided. 



\begin{algorithm}  
\caption{System-AAA}
\label{sys_aaa_alg}
\begin{algorithmic} 
        \Require \(G(s)\) in state space form
        \State \(k \gets 0\)
        \State \(R \gets \mathrm{ss}(G_D)\)
        \State \(NM \gets \mathrm{ss}()\)
        \Repeat
                \State \(\omega_k \gets \mathrm{hinfnorm}(G-R)\)
                \State \(G_k = G(\omega_i), \; G_{k, r} = \mathrm{real}(G_k), \; G_{k, i} = \mathrm{imag}(G_k)\)
                \If{\(\omega_k = 0\)}
                        \State \(NM_k \gets \brac{\begin{array}{c|cc}0 & G_i & I \\ \hline I & 0 & 0\end{array}}\)
                \Else 
                        \State \(NM_k \gets \brac{\begin{array}{cc|cc}0 & \omega_k I & G_{k, r} & I \\ -\omega_k I & 0 & -G_{k, i} & 0 \\ \hline I & 0 & 0 & 0 \\ 0 & I & 0 & 0 \end{array}}\)
                \EndIf
                \State \(NM \gets \begin{bsmallmatrix}NM \\ NM_k\end{bsmallmatrix}\)
                \State \(H \gets \mathrm{minreal} \p{NM \begin{bsmallmatrix}I \\ -G\end{bsmallmatrix}}\) 
                \State \(X \gets H_C \p{\mathrm{lyap}(H_A, H_BH_B^*)} H_C^*\)
                \State Construct \(\mbb{W}\) using theorem \ref{opt_thm}
                \State \(\mcal{B}_1 \gets NM_B(:, 1:m)\)
                \State \(\mcal{B}_2 \gets NM_B(:, m+1:\mathrm{end})\)
                \State \(R \gets \brac{\begin{array}{c|c}NM_A - \mcal{B}_2 \widehat{\mbb{W}} & \mcal{B}_2 G_D - \mcal{B}_1 \\ \hline -\widehat{\mbb{W}} & G_D\end{array}}\)
                \State \(i \gets i + 1\)
        \Until{\(R\) approximates \(G\) sufficiently}
        \State \textbf{return} \(R\)
\end{algorithmic}
\end{algorithm}
Algorithm \ref{sys_aaa_alg} loosely follows MATLAB notation.  The subscripts \(A\), \(B\), \(C\), and \(D\) denote the corresponding state-space matrix for the system.  The following subsections detail the derivation of the algorithm and its connections to AAA.  The first subsection uses the block-AAA interpolating function as a basis and derives a new interpolating function that interpolates at \(\omega=j\infty\), guarantees real coefficients, and derives its associated transfer functions.  The second subsection details the transformation of the block-AAA algorithm into a state-space context.  The third details the derivation of the state-space representation of the interpolation function.  Finally the final section shows computational results for the algorithm.  

\subsection{Interpolation function}
Consider the multi-input, multi-output (MIMO) system 
\[G = \brac{\begin{array}{c|c}A & B \\ \hline C & D\end{array}},\]
where \(A\in\real^{n\times n}\), \(B\in\real^{n\times q}\), \(C\in\real^{p\times n}\), and \(D\in\real^{p\times q}\). We choose support points that always lie on the imaginary axis, thus Equation~(\ref{blockAAA}) becomes
\begin{equation}
        R_r(s) = \p{\sum_{k=1}^r \frac{W_k}{s - j\omega_k}}\inv \p{\sum_{k=1}^r \frac{W_k G(j\omega_k)}{s-j\omega_k}}. \label{rr_eq}
\end{equation}
\begin{remark} \label{rem_int}
        The interpolating function (\ref{rr_eq}) guarantees that \(R_r(j\omega_i) = G(j\omega_i)\) for any support point \(\omega_i\), \(1\leq i \leq r\), provided that each \(W_k\) is invertible.    

        \begin{proof}
                Multiplying by \(\frac{s-j\omega_i}{s-j\omega_i}\) yields 
                \begin{align*}
                    R_r(s) =& \p{W_i + \sum_{i\neq k = 1}^r \frac{(s-j\omega_i) W_k}{s-j\omega_k}}\inv \\
                    & \p{W_iG(j\omega_i) + \sum_{i\neq k = 1}^r \frac{(s-j\omega_i) W_kG(j\omega_k)}{s-j\omega_k}} \\
                    \therefore R_r(j\omega_i) =& W_i\inv W_i G(j\omega_i) = G(j\omega_i).
                \end{align*}
        \end{proof}
\end{remark}

From here, we begin to address the issues that were outlined above.  The algorithm needs the ability to interpolate at \(\omega=j\infty\).  We therefore rewrite the interpolation in a more general form, yielding
\begin{subequations} \label{rmn_eq}
        \begin{align}
                R_\ell(s) &= M(s)\inv N(s), \\
                \shortintertext{where}
                M(s) &= \mcal{W}_0 + \sum_{k=1}^\ell \mcal{W}_k M_k(s) \\
                N(s) &= \mcal{W}_0 D + \sum_{k=1}^\ell \mcal{W}_k N_k(s), 
        \end{align}
\end{subequations}
and \(M(s)\in\comp^{p\times p}\), \(N(s)\in\comp^{p\times q}\).  All of the weights in \(M(s)\) and \(N(s)\) can be factored out to the left, meaning \(M\) and \(N\) can be written as
\begin{align}
    N(s) &= \mbb{W}\mcal{N}(s), \quad M(s) = \mbb{W}\mcal{M}(s), \non \\
    \shortintertext{where}
    \mbb{W} &= \begin{bmatrix}\mcal{W}_0 & \mcal{W}_1 & \cdots & \mcal{W}_\ell\end{bmatrix} \non \\
    \mcal{N}(s) &= \begin{bsmallmatrix}D \\ N_1(s) \\ \vdots \\ N_\ell(s)\end{bsmallmatrix}, \qquad \mcal{M}(s) = \begin{bsmallmatrix}I \\ M_1(s) \\ \vdots \\ M_\ell(s)\end{bsmallmatrix}. \non 
\end{align}
Note \(\mbb{W}\in\real^{p\times pr}\).  Depending on the location of the support point, the size of \(M_k(s)\) or \(N_k(s)\) can change.  To ensure the resulting interpolating function has real coefficients, it must be the case that \(G(j\omega) = \overline{G}(-j\omega)\) for any \(\omega\in\real\).  This may be accomplished by adding pairs of complex conjugate support points with conjugate weights.  Starting with \(M\),
\begin{align}
    \mcal{W}_kM_k(s) &= \frac{W_{k,1}}{s-j\omega_k} + \frac{W_{k, 2}}{s+j\omega_k}. \non \\
    \shortintertext{Assuming \(W_{k,1} = W_k\) and \(W_{k, 2} = \overline{W}_k\),}
    \mcal{W}_kM_k(s) &= \frac{2s\Re(W_k)-2\omega_k\Im(W_k)}{s^2 + \omega_k^2}. \non \\
    \shortintertext{Therefore}
    \mcal{W}_k &= 2\begin{bmatrix}\Re(W_k) & \Im(W_k)\end{bmatrix} \non \\ 
    M_k(s) &= \begin{bmatrix}\frac{s}{s^2 + \omega_k^2} \\ -\frac{\omega_k}{s^2 + \omega_k^2}\end{bmatrix}. \label{mk_eq} \\
    \shortintertext{Similarly for \(N\),}
    \mcal{W}_k N_k(s) &= \frac{W_{k, 1}G(j\omega_k)}{s-j\omega_k} + \frac{W_{k, 2}\overline{G}(j\omega_k)}{s+j\omega_k} \non \\
    \therefore N_k(s) &= \begin{bmatrix}\frac{\Re(G(j\omega_k))s - \Im(G(j\omega_k))\omega_k}{s^2+\omega_k^2} \\ -\frac{\Im(G(j\omega_k))s + \Re(G(j\omega_k)\omega_k}{s^2+\omega_k^2} \end{bmatrix}. \label{nk_eq} 
\end{align} 
In this case \(\mcal{W}_k\in\real^{p\times 2p}\), \(M_k(s)\in\comp^{2p\times p}\), and \(N_k(s)\in\comp^{2p\times q}\). When \(\omega_k=0\), the first order system is already real thus there is no need to add an additional complex conjugate support point.  In this case,
\begin{equation}
        M_k(s) = \frac{I}{s}, \quad N_k(s) = \frac{G(0)}{s}, \quad \mcal{W}_k = W_k, \label{nm0_eq}
\end{equation}
and \(\mcal{W}_k\in\real^{p\times p}\), \(M_k(s)\in\comp^{p\times p}\), and \(N_k(s)\in\comp^{p\times q}\).

\subsection{Algorithm reformulation}
Each step of the AAA algorithm is composed of two main parts, the first being the selection of the new support point at the beginning of each iteration. The second is the selection of the weight matrices from an optimization problem that minimizes the approximation error between the interpolating function and the input function.  In this section we show that these parts can be reformulated remove the necessity of a user-defined domain and to better utilize systems machinery.  

The next support point is chosen at the point in the domain where the error between \(R_r(z)\) and \(G(z)\) is largest.  The domain in this case is the imaginary line, so the next support point will be at the frequency where the \(\mcal{H}_\infty\) norm occurs, i.e.
\begin{equation}
        \omega_\ell = \arg\min_{\omega\in\real_{\geq 0}} \norm{G(j\omega) - R_{\ell-1}(j\omega)}_2.
\end{equation}
This can be efficiently calculated with a bisection algorithm \cite{Bruinsma90}.  After a support point is selected, the weights in the interpolating function are selected via an optimization problem.  The optimization problem in block-AAA is the following: 
\[\min_{\mbb{W}} \sum_{z\in\Omega} \norm{N(z)-M(z)G(z)}^2_F \quad \st \norm{\mbb{W}}_F = 1.\]
Since our analysis in in continuous time, the sum will be replaced with an integral over the positive imaginary axis yielding
\begin{align}
    \mbb{W} &= \arg\min_{\mbb{W}} \int_0^\infty \norm{\mbb{W}\p{\mcal{N}(j\omega)-\mcal{M}(j\omega)G(j\omega)}}_F^2 \rmd\omega. \non \\
    \shortintertext{Letting \(H(s) = \mcal{N}(s) - \mcal{M}(s)G(s)\),}
    &= \arg\min_{\mbb{W}} \int_0^\infty \tr\p{\mbb{W}H(j\omega)H^*(j\omega)\mbb{W}^*} \rmd\omega \non \\
    &= \arg\min_{\mbb{W}} \tr\p{\mbb{W}X\mbb{W}^*}, \non \\
    \shortintertext{where}
    X &= \int_0^\infty H(j\omega)H^*(j\omega)\rmd\omega = \hat{C}\hat{G}_C\hat{C}^*, \label{x_eq}
\end{align}
where \(G_C\) is the controllability Gramian for \(H\).  Note that \(H\) can be written as a product of two augmented systems,
\begin{equation}
        H = \begin{bmatrix}\mcal{N} & \mcal{M}\end{bmatrix} \begin{bmatrix}I \\ -G\end{bmatrix}, \label{h_eq}
\end{equation}
and the positive matrix \(G_C\) can be found via the Lyapunov equation \cite[p. 112]{Zhou95} 
\begin{equation}
        \hat{A}G_C + G_C\hat{A}\trans = -\hat{B}\hat{B}^*, \label{x_lyap}
\end{equation}
where \(\hat{A}\), \(\hat{B}\), and \(\hat{C}\) are the corresponding state space matrices of \(H\).
\begin{remark}
        In order to guarantee existence and uniqueness of \(G_c\), the system \(H\) must not have any marginally stable poles, i.e. \(\hat{A}\) must not have any eigenvalues on the imaginary axis.  However, this system has poles at \(\pm j\omega_k\) for all support points \(\omega_k\).  It can be shown that there is a pole-zero cancellation for all of these poles, thus finding a minimal realization of \(H\) will suffice to find \(G_c\).  
\end{remark}
The constraint \(\norm{\mathbb{W}}_F = 1\) is modified to \(\mathbb{W}\mathbb{W}^* = I\) in the new problem to guarantee \(\mbb{W}\) has full row rank.  Therefore the optimization becomes:
\begin{equation}
        \mbb{W} = \arg\min_{\mbb{W}} \tr\p{\mbb{W}X\mbb{W}^*}, \quad \st \mbb{W}\mbb{W}^* = I. \label{opt_eq} 
\end{equation}
The closed form for (\ref{opt_eq}) may be found by finding stationary points.  A necessary condition for optimality is the following:
\begin{equation}
        \mbb{W}X - \Lambda^* \mbb{W} = 0. \label{eig_eq}
\end{equation}
\begin{theorem} \label{opt_thm}
        A solution for equation (\ref{eig_eq}) subject to \(\mbb{W}\mbb{W}^* = I\) is \(\mathbb{W} = QV^*\), where \(Q\) is an arbitrary real unitary matrix and the columns of \(V\) are the eigenvectors corresponding to the \(p\) smallest distinct non-zero eigenvalues of \(X\).  

        \begin{proof}
                Rearranging (\ref{eig_eq}) yields \(X\mbb{W}^* = \mbb{W}^*\Lambda\).  This implies \(\mbb{W}X\mbb{W}^* = \mbb{W}\mbb{W}^*\Lambda = \Lambda\).  Taking the conjugate transpose shows that \(\Lambda = \Lambda^*\), therefore \(\Lambda\) can be diagonalized.  Letting \(\Lambda = UDU^*\) where \(U\) is unitary and \(D\) is a diagonal matrix yields
                \[X\mbb{W}^* = \mbb{W}^* X \implies XV = VD,\]
                where \(V = \mbb{W}^* U\).  Now let \(V = \begin{bsmallmatrix}v_1 & \cdots & v_p\end{bsmallmatrix}\) and \(D = \diag\curly{\begin{bsmallmatrix}d_1 & \cdots & d_p\end{bsmallmatrix}}\).  Then,
                \[Xv_1 = d_1 v_1, \quad \cdots \quad Xv_p = d_p v_p,\]
                This shows that the columns of \(V\) are eigenvectors of \(X\) and the diagonal elements of \(D\) are the corresponding eigenvalues.  
                
                From \(V=\mbb{W}^*U\), clearly \(\mbb{W} = UV^*\), therefore \(\mbb{W}\mbb{W}^* = UV^*VU^* = V^*V = I\).  This demonstrates the columns of \(V\) must be orthonormal, which implies the eigenvectors that are picked must be associated with distinct eigenvalues.  
                
                Recall the optimization problem is \(\min\tr\p{\mbb{W}X\mbb{W}^*} = \min\tr\p{\mbb{W}\mbb{W}^*\Lambda} = \min\tr\Lambda = \min\tr{UDU^*} = \min\tr D\).  This implies the eigenvectors we pick must correspond to the smallest eigenvalues.    
        \end{proof}
\end{theorem}
\subsection{State-space realizations}
In this subsection we show that there exists a state-space representation for the interpolation function.  Equation (\ref{h_eq}) requires a realization for \(\begin{bsmallmatrix}\mcal{N} & \mcal{M}\end{bsmallmatrix}\).  This system is a vertical concatenation of the individual \(\begin{bsmallmatrix}N_k & M_k\end{bsmallmatrix}\) systems.  We can create a state space representation of this combined system by considering equations (\ref{mk_eq}), (\ref{nk_eq}), and (\ref{nm0_eq}).  For \(\omega_k = 0\), clearly
\begin{equation}
        \begin{bmatrix}N_k & M_k\end{bmatrix} = \brac{\begin{array}{c|cc}0 & G_k & I \\ \hline I & 0 & 0\end{array}}.
\end{equation}
When \(\omega_k \neq 0\), the equations are more complicated, so the realization is slightly harder to derive:
\begin{equation}
        \begin{bmatrix}N_k & M_k\end{bmatrix} = \brac{\begin{array}{cc|cc}0 & \omega_k I & G_{k, r} & I \\ -\omega_k I & 0 & -G_{k, i} & 0 \\ \hline I & 0 & 0 & 0 \\ 0 & I & 0 & 0 \end{array}},
\end{equation}
where \(G(j\omega_k) = G_k\), \(\Im(G_k) = G_{k, i}\), and \(\Re(G_k) = G_{k, r}\).  In general, this is written as 
\begin{equation}
        \begin{bmatrix}N_k & M_k\end{bmatrix} = \brac{\begin{array}{c|cc}A_k & B_{k, 1} & B_{k, 2} \\ \hline I & 0 & 0\end{array}}. \label{nkmk_eq}
\end{equation}
Notice that in either case, the size of \(A_k\) is a multiple of the number of outputs \(p\).  Now let \(\mathcal{A}\) be a block diagonal matrix composed of \(A_k\) matrices, and \(\mcal{B}_1\) and \(\mcal{B}_2\) be the block column vectors of \(B_{k, 1}\) and \(B_{k, 2}\) matrices respectively. This leads to the state-space representation of the interpolating function \(R_\ell\).  
\begin{lemma}
        Let 
        \begin{equation}
                N = \brac{\begin{array}{c|c}\mcal{A} & \mcal{B}_1 \\ \hline \mbb{W}_1 & \mcal{W}_0 D\end{array}}, \; M = \brac{\begin{array}{c|c}\mcal{A} & \mcal{B}_2 \\ \hline \mbb{W}_1 & \mcal{W}_0\end{array}}. \label{n_m_eq}
        \end{equation}
        where \(\mbb{W} = \begin{bsmallmatrix}\mcal{W}_0 & \mbb{W}_1\end{bsmallmatrix}\).  Assuming \(\mcal{W}_0\) is invertible,
        \begin{equation}
                R_\ell = M\inv N = \brac{\begin{array}{c|c}\mcal{A}-\mcal{B}_2 \W & \mcal{B}_2 D - \mcal{B}_1 \\ \hline -\W & D \end{array}}. \label{rss_eq}
        \end{equation}

\begin{proof}
Vertically concatenating each \(\begin{bsmallmatrix}N_k & M_k\end{bsmallmatrix}\) to create \(\begin{bsmallmatrix}\mcal{N} & \mcal{M}\end{bsmallmatrix}\) yields
\begin{equation}
        \begin{bmatrix}\mcal{N} & \mcal{M}\end{bmatrix} = \brac{\begin{array}{c|cc}\mcal{A} & \mcal{B}_1 & \mcal{B}_2 \\ \hline 0 & D & I \\ I & 0 & 0\end{array}}. 
\end{equation}    
Multiplying by \(\mbb{W}\) and separating \(N\) and \(M\) yields equation (\ref{n_m_eq}).  Then, 
\[M\inv = \brac{\begin{array}{c|c}\mcal{A}-\mcal{B}_2 \mcal{W}_0\inv \mbb{W}_1 & \mcal{B}_2 \mcal{W}_0\inv \\ \hline -\mcal{W}_0\inv \mbb{W}_1 & \mcal{W}_0\inv\end{array}}.\]
This shows 
\[M\inv N = \brac{\begin{array}{cc|c}\mcal{A}-\mcal{B}_2 \W & \mcal{B}_2 \W & \mcal{B}_2 D \\ 0 & \mcal{A} & \mcal{B}_1 \\ \hline -\W & \W & D\end{array}},\]
where \(\W = \mcal{W}_0\inv \mbb{W}_1\).  Applying a coordinate change with transformation matrix \(\frac{1}{2}\begin{bsmallmatrix}I & I \\ -I & I\end{bsmallmatrix}\), we get 
\[M\inv N = \brac{\begin{array}{cc|c}\mcal{A}-\mcal{B}_2 \W & 0 & \mcal{B}_2 D - \mcal{B}_1 \\ -\mcal{B}_2 \W & \mcal{A} & \mcal{B}_2 D + \mcal{B}_1 \\ \hline -\W & 0 & D\end{array}}.\]
From this it is clear the second block mode is unobservable since it affects neither the output nor the first block state, therefore it may be eliminated to yield the final representation for the approximating system.  
\end{proof}
\end{lemma}

\begin{remark} \label{rem_p}
        One \(A_k\) matrix is appended to \(\mcal{A}\) at every iteration, showing that the system grows by a multiple of \(p\) states each time. 
\end{remark}

\subsection{Computational results}
\begin{figure}[!ht]
        \includegraphics[width=\linewidth]{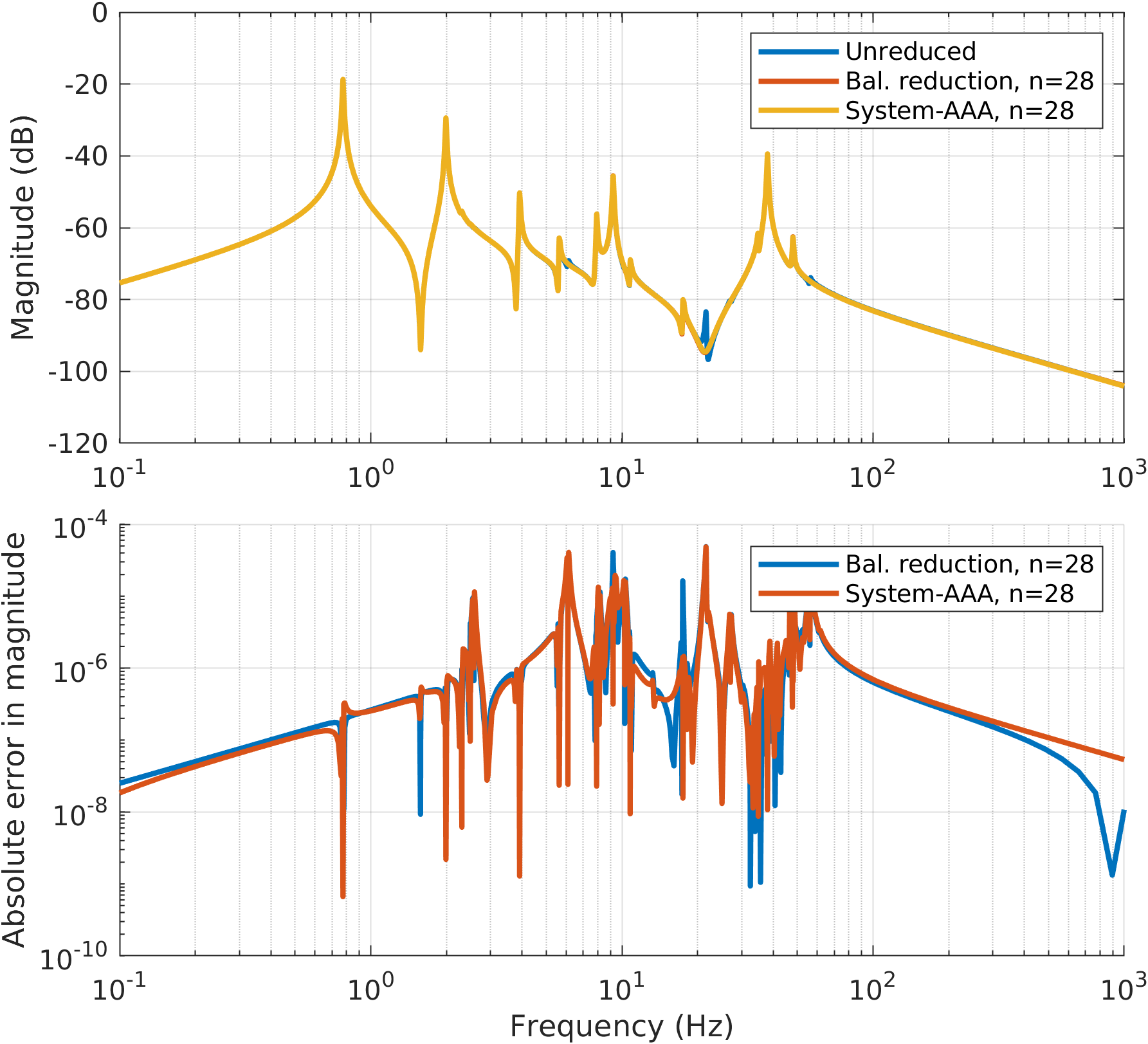}
        \caption{ISS single-output \(n=28\) reduction}
        \label{cont_iss11_n27}
\end{figure}

\begin{figure}[!ht]
        \includegraphics[width=\linewidth]{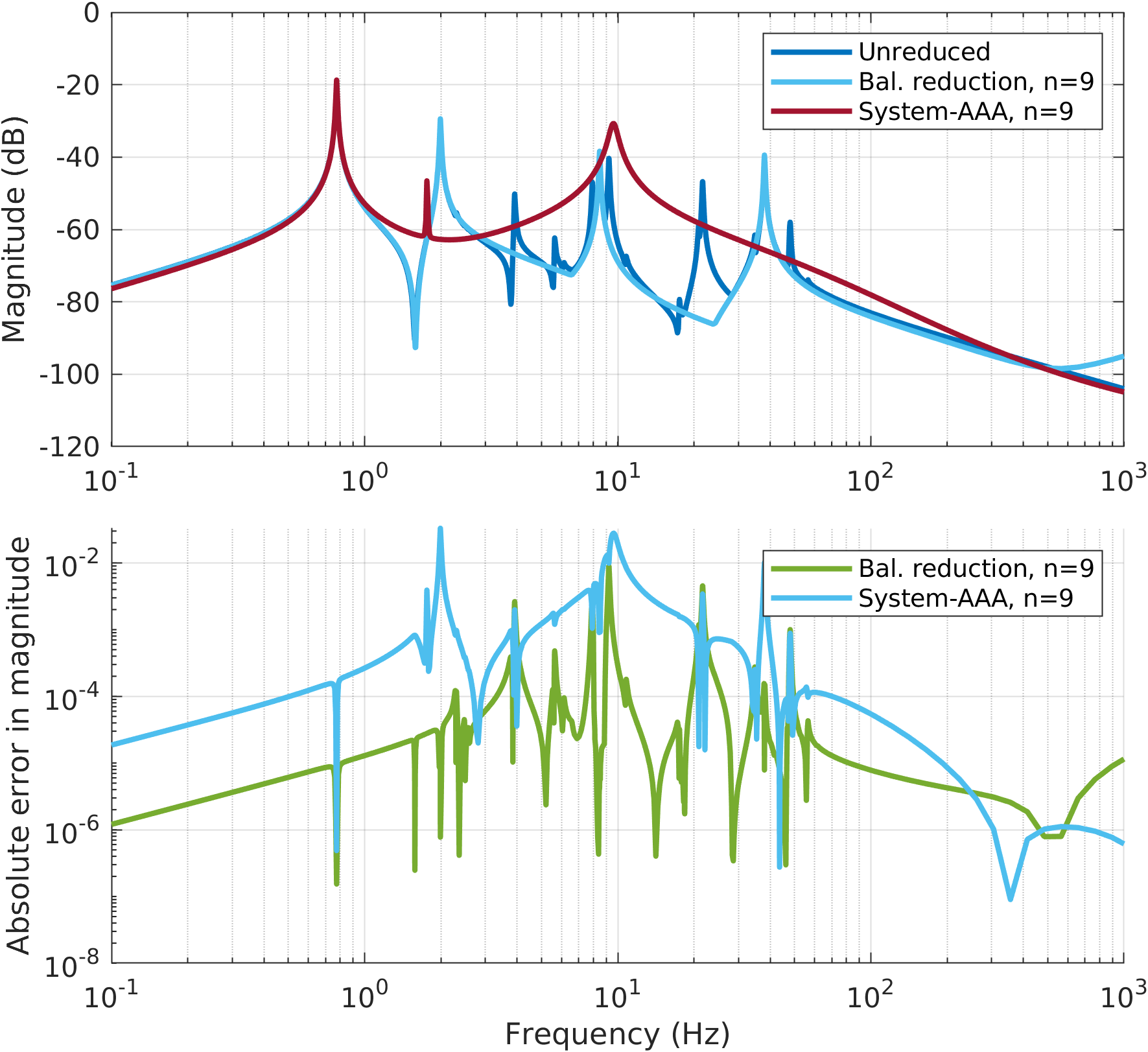}
        \caption{ISS \(n=9\) reduction}
        \label{cont_iss_n9}
\end{figure}

In this section, we discuss some numerical results where we compare our algorithm with a baseline, i.e. balanced reduction.  Balanced reduction is a standard algorithm for model reduction on LTI systems~\cite{Zhou95}.  We use as a test case a 270-state, 3-input, 3-output stable dynamic system modeling the dynamics of a module on the International Space Station (ISS) \cite{iss_model}.   The figures each show a plot of the maximum singular value of the frequency response for the reduced-order systems, and the absolute error between the reduced order systems and the full system.  The model used in figure \ref{cont_iss11_n27} is a 28-state approximation of the ISS system's first output only, while figure \ref{cont_iss_n9} shows the approximation on the full system.  The ISS system was reduced using both system-AAA and balanced reduction, and the figures demonstrate the difference in approximation error. 

Figure \ref{cont_iss11_n27} shows that the algorithm generated a stable and well-matched approximation to the system with comparable error to that of standard balanced reduction when used on a single-output system.

Figure \ref{cont_iss_n9} shows the result when the algorithm is used on a MIMO system after two iterations.  In this case, the algorithm selected two support points at \(\omega=0\) and \(\omega\approx 0.8\) Hz.  Balanced reduction is able to stably replicate the dynamics at 4 peaks, while system-AAA only mirrors one peak and has unstable poles.  This effect becomes more pronounced as more outputs are added.  As stated in remark \ref{rem_p}, the number of states added per iteration is proportional to the number of outputs, suggesting that an improved algorithm would not have this dependence.  As mentioned in remark \ref{rem_int}, invertible \(W_k\) matrices are required for interpolation.  Numerous numerical simulations demonstrate that these \(W_k\) matrices are well-conditioned.   These observations motivate us to propose a different algorithm as stated in the next section.

\section{Low-rank approximation}
Though the performance of system-AAA is satisfactory with single-output systems, the results indicate that the performance degrades as the number of outputs increases.  In order to rectify this, we investigated a slight change to remove the system's size dependence on the number of outputs with an algorithm we shall call low-rank approximation.  

\subsection{Interpolation function}
With low-rank approximation, we allow the approximating system to be non-full-rank at the support points.  This ensures that the interpolation function will grow one state when a new support point is added.  Consider the following approximating function
\begin{equation}
        R_r(s) = \p{\sum_{k=1}^r \frac{W_k U_k^*}{s-j\omega_k}}^\dagger \p{\sum_{k=1}^r \frac{W_k \Sigma_k V_k^*}{s-j\omega_k}}, \label{pir_eq}
\end{equation}
where \(U_k\Sigma_k V_k^* = G(j\omega_k)\) is a rank \(r_k\) approximation.  Let \(U_k\in\comp^{p\times r_k}\), \(V\in\comp^{m\times r_k}\), and \(\Sigma_k\in\real^{r_k \times r_k}\), and assume \(U_k\) and \(V_k\) have orthonormal columns, and \(\Sigma_k\) is diagonal.   When \(r_k=1\), \(U_k\), \(V_k\), \(\Sigma_k\), and \(W_k\) are all rank 1, showing that this approximation function will clearly not fully interpolate at the support point.   When \(r_k=p\), the approximation is full rank and fully interpolates the corresponding support point.   

Akin to system-AAA, \(R_r(s)\) can be rewritten to yield a \(M(s)\) and \(N(s)\) that are in the same form as (\ref{rmn_eq}).  Though \(R_r(s)\) contains a pseudoinverse, the system inverse \(M\inv\) is well-defined as long as \(\mcal{W}_0\) is invertible like before, thus the pseudoinverse is replaced with \(M\inv\).  Note that \(U_k \in\comp^{p\times r_k}\) and \(W_k\in\comp^{p\times r_k}\), so their product can only be full rank when \(r_k = p\).  
\begin{remark}
If \(p>q\), then when \(r_k = p\), \(V_k\) is not full rank, so the resulting \(R_\ell(j\omega_k)\) will not be full rank.  Therefore, we may perform model reduction on the dual of \(G(s)\), i.e. 
\[\mathrm{dual}[G](s) := \brac{\begin{array}{c|c}A\trans & C\trans \\ \hline B\trans & D\trans\end{array}}.\]
After the model is satisfactory, then we may return the dual of the reduced system.  
\end{remark}

The transfer functions of \(M_k\) and \(N_k\) are similar to the forms seen in the full interpolation algorithm, except the \(M_k\) systems have an added matrix \(U_k^*\), making their transfer functions more similar to \(N_k\).  The form of \(M_k\) and \(N_k\) are the following when \(\omega_k = 0\),
\[M_k(s) = \frac{U_k\trans}{s}, \quad N_k(s) = \frac{\Sigma_k V_k\trans}{s}, \quad \mcal{W}_k = W_k,\]
and when \(\omega_k \neq 0\),
\begin{align*}
    M_k(s) &= \begin{bmatrix}\frac{U_{k, r}\trans s + U_{k, i}\trans\omega_k}{s^2 + \omega_k^2} \\ \frac{U_{k, i}\trans s - U_{k, r}\trans\omega_k}{s^2 + \omega_k^2}\end{bmatrix} \\
    N_k(s) &= \begin{bmatrix}\frac{\Sigma_k V_{k, r}\trans s + \Sigma_k V_{k, i}\trans\omega_k}{s^2 + \omega_k^2} \\ \frac{\Sigma_k V_{k, i}\trans s - \Sigma_k V_{k, r}\trans\omega_k}{s^2 + \omega_k^2}\end{bmatrix}.
\end{align*}
The state space realizations for \(\begin{bsmallmatrix}N_k & M_k\end{bsmallmatrix}\) for \(\omega_k = 0\) is 
\begin{align*}
    \begin{bmatrix}N_k & M_k\end{bmatrix} &= \brac{\begin{array}{c|cc}0 & \Sigma_k V_k\trans & U_k\trans \\ \hline I & 0 & 0\end{array}}, \\
    \shortintertext{and for \(\omega_k \neq 0\),}
    \begin{bmatrix}N_k & M_k\end{bmatrix} &= \brac{\begin{array}{cc|cc}0 & \omega_k I & \Sigma_k V_{k, r}\trans & U_{k, r}\trans \\ -\omega_k I & 0 & \Sigma_k V_{k, i}\trans & U_{k, i}\trans \\ \hline I & 0 & 0 & 0 \\ 0 & I & 0 & 0 \trans\end{array}}.
\end{align*}
Note that \(U_{k, r}\), \(U_{k, i}\) are the real and imaginary parts of \(U_k\) respectively, and similarly for \(V_{k, r}\) and \(V_{k, i}\).  

\subsection{Algorithm}
The main change between system-AAA and the low-rank approximation algorithm is the modification of the approximating function.  This does not affect the majority of the algorithm.  However, the rank of the approximation at each support point needs to be addressed.  When a new support point is added, it will always start out as a rank 1 approximation, but the algorithm must also consider whether the improvement of the approximation at an existing support point will be more effective.  To do this, after a candidate \(\omega_\ell\) is selected, it will be compared to the previous support points, and if it is close to an existing support point, then it will instead improve said support point.  The minimum distance to a support point then is a tunable parameter.  

\subsection{Computational Results}
The ISS model was used again as a test for the partial approximation algorithm.  Like before, the following figures show the maximum singular value plot and its absolute error for a various number of states in each reduced system.  

\begin{figure}[!ht]
        \includegraphics[width=\linewidth]{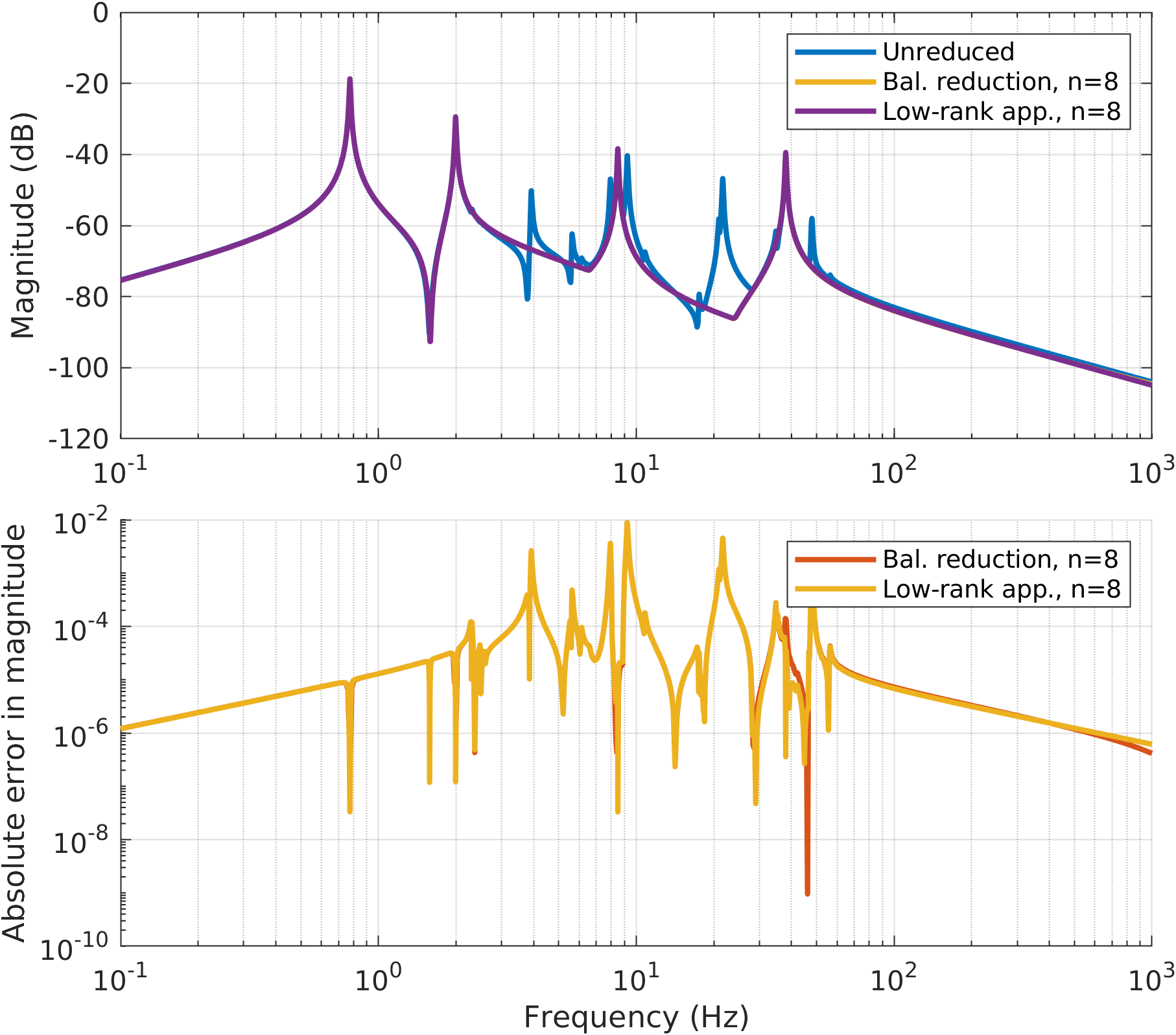}
        \caption{ISS \(n=8\) reduction}
        \label{part_iss_n8}
\end{figure}

\begin{figure}[!ht]
        \includegraphics[width=\linewidth]{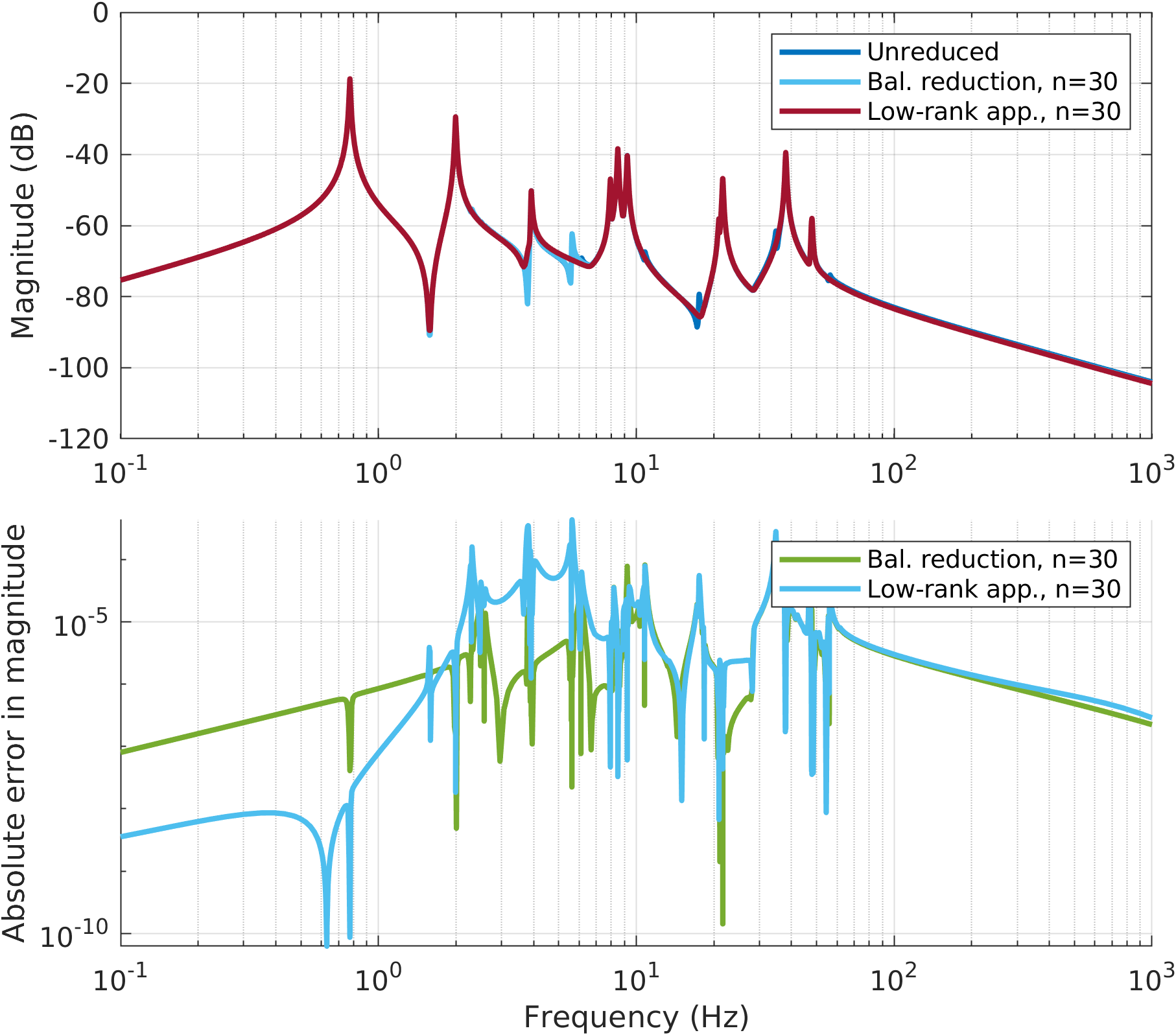}
        \caption{ISS \(n=30\) reduction}
        \label{part_iss_n30}
\end{figure}

Figures \ref{part_iss_n8} and \ref{part_iss_n30} show the results for the low-rank approximation algorithm.  From here it is clear the dynamics at more peaks are being incorporated compared to full interpolation.  In general, the results outperform full interpolation and are close to that of balanced reduction.  

\begin{figure}[ht!]
        \includegraphics[width=\linewidth]{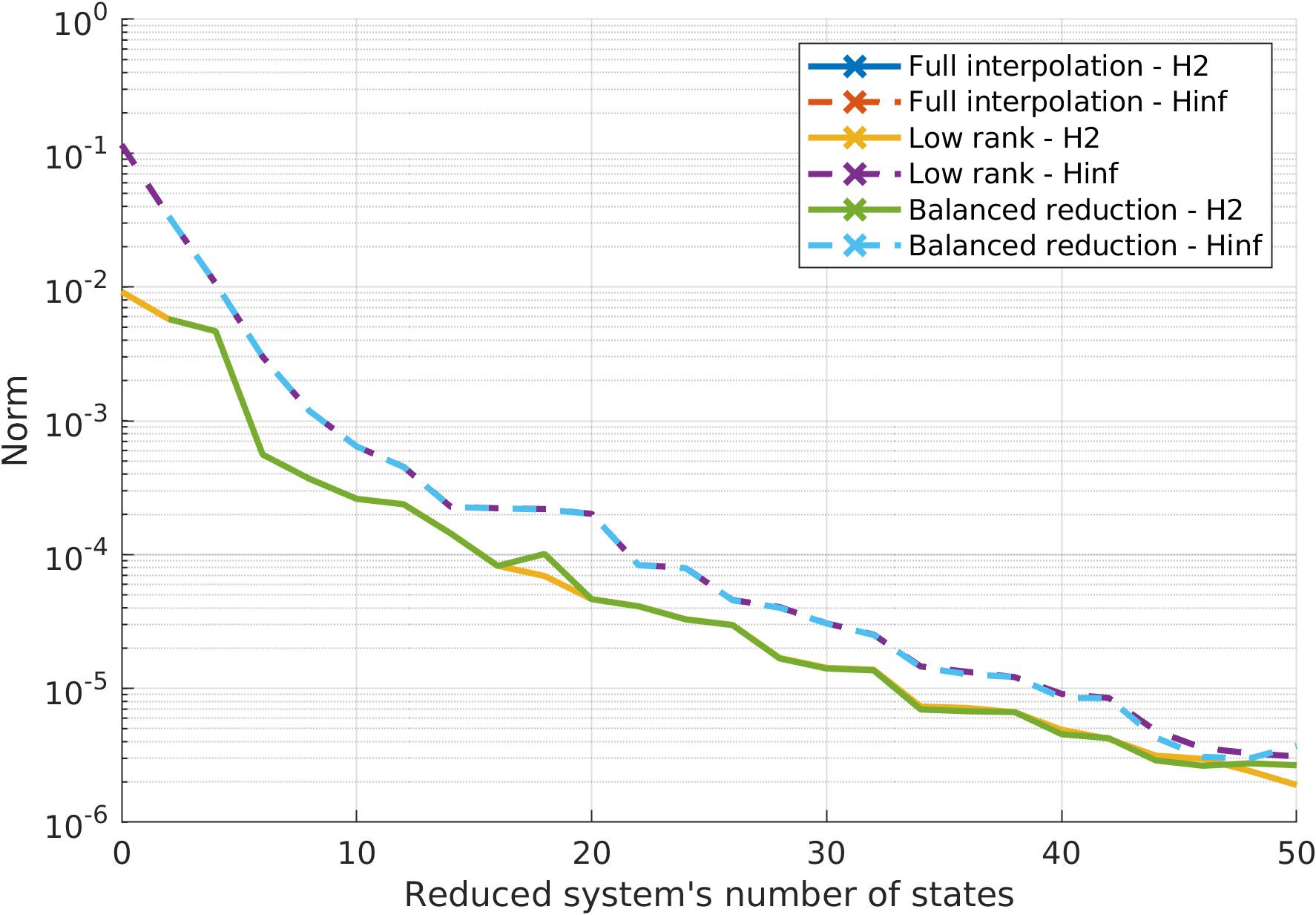}
        \caption{SISO system error as number of states increases}
        \label{error_siso}
\end{figure}

\begin{figure}[ht!]
        \includegraphics[width=\linewidth]{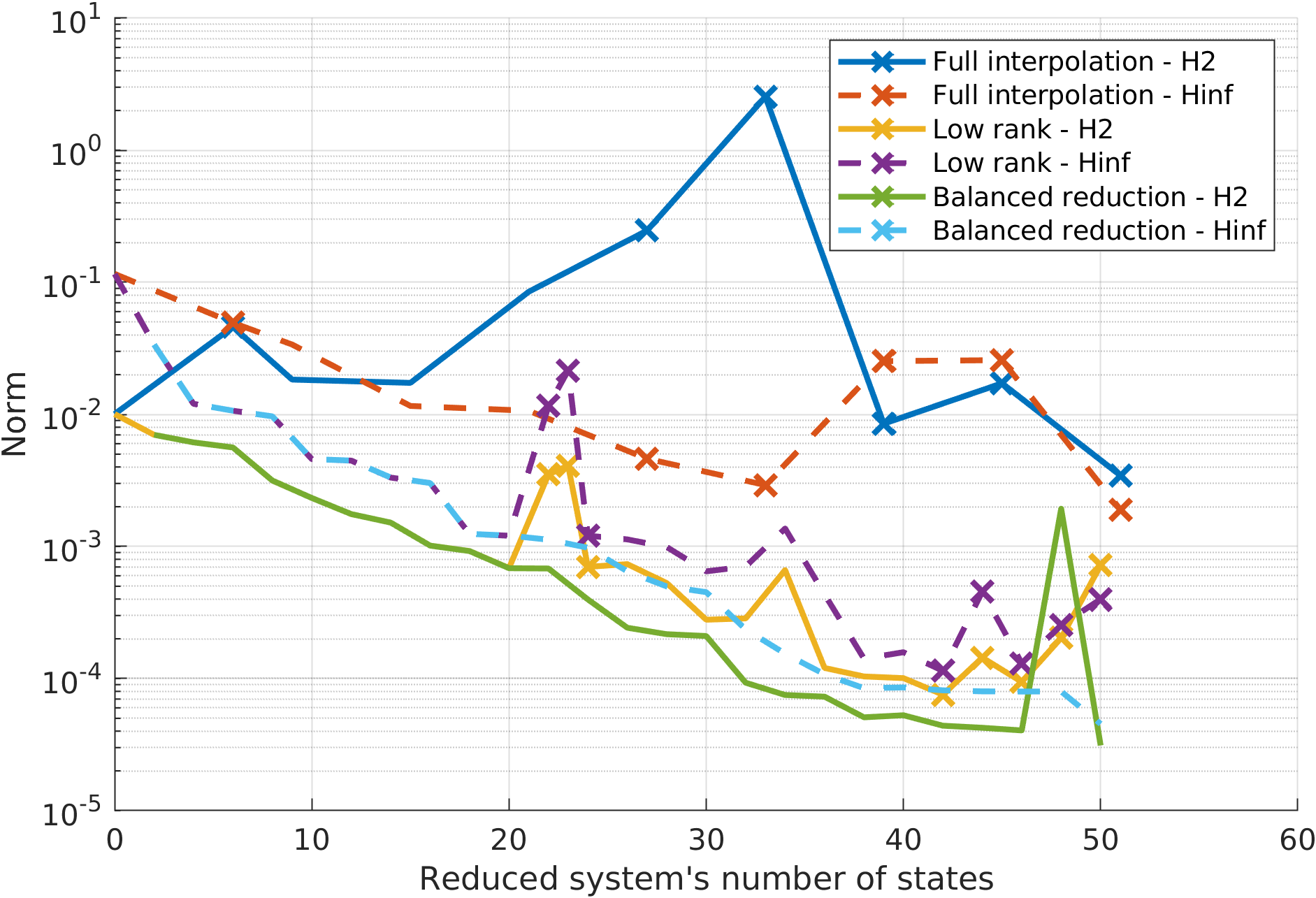}
        \caption{MIMO system error as number of states increases}
        \label{error_mimo}
\end{figure}

Figures \ref{error_siso} and \ref{error_mimo} show the approximation error as the number of states increases for the two algorithms presented in this paper as well as balanced reduction.  The \(\mcal{H}_\infty\) norm indicated is the maximum error over the frequency domain, and the \(\mcal{H}_2\) norm written is the error integrated across the domain.  More precisely, in this context it has been calculated as:
\[\sqrt{\left|\tr\p{\hat{C}P\hat{C}^*}\right|}, \quad \hat{A}P + P\hat{A}^* = \hat{B}\hat{B}^*, \]
where \(\hat{A}\), \(\hat{B}\), and \(\hat{C}\) are the corresponding state space matrices to \(G-R_\ell\), the system representing the error between the input system and the reduced order system.  The presence of an `x' indicates that the resulting reduced order system had unstable poles with that number of states.  The first figure shows the error for the (1, 1) channel of the ISS system, while the second figure shows the error for the entire ISS system.  

It is clear to see that in the SISO and MISO case, both algorithms perform well and match the performance of balanced reduction, yielding a stable reduced system.  For MIMO systems, the results are much more interesting and indicate a few things.  The error for full interpolation may increase as the number of states increases, and doesn't always yield a great result for some number of states.  In addition to this, most of the resulting systems contain a number of unstable poles.  

In comparison to these observations, Low-rank approximation matches the performace of balanced reduction up until a certain number of states, at which point the error slightly increases.  Low-rank approximation may generate systems with a few unstable poles, but does not always, indicating that the user may stop the algorithm once a satisfactorily-performing stable system is found.  Overall, the low rank approximation algorithm gives better results compared to full interpolation, and can give comparable results to balanced reduction.  

\section{Discussion}
In this paper, we adapted the AAA algorithm for use in the model order reduction of state space systems.  The first algorithm, system-AAA, gives satisfactory results for single-output systems, but does not perform as strongly when compared to balanced reduction with multi-output systems.  We also discussed a second algorithm, low-rank approximation, which removes the state dimension's dependence on the number of outputs.  Low-rank approximation fixes some issues with full interpolation and yields improved results with MIMO systems.  Numerical results show that this new algorithm performs similarly to balanced reduction with MIMO systems, and matches or exceeds its performace otherwise.  For single-output systems, both system-AAA and low-rank approximation are good alternatives to balanced reduction when the user needs a minimum order system.  Starting with a minimum order system and gradually increasing the order allows the user to choose the smallest system that meets certain \(\mcal{H}_\infty\) or \(\mcal{H}_2\) error requirements, which is an advantage over other model reduction techniques.  

In future work, we will investigate why both algorithms can produce unstable poles in the MIMO case.  We would like to find ways to further improve the performace of low-rank system-AAA, namely by ensuring the algorithm yields a stable, well-matched result on MIMO systems.  

\bibliographystyle{IEEEtran}
\bibliography{IEEEabrv,paper_bib}

\begin{thebibliography}{10}
\providecommand{\url}[1]{#1}
\csname url@rmstyle\endcsname
\providecommand{\newblock}{\relax}
\providecommand{\bibinfo}[2]{#2}
\providecommand\BIBentrySTDinterwordspacing{\spaceskip=0pt\relax}
\providecommand\BIBentryALTinterwordstretchfactor{4}
\providecommand\BIBentryALTinterwordspacing{\spaceskip=\fontdimen2\font plus
\BIBentryALTinterwordstretchfactor\fontdimen3\font minus
  \fontdimen4\font\relax}
\providecommand\BIBforeignlanguage[2]{{%
\expandafter\ifx\csname l@#1\endcsname\relax
\typeout{** WARNING: IEEEtran.bst: No hyphenation pattern has been}%
\typeout{** loaded for the language `#1'. Using the pattern for}%
\typeout{** the default language instead.}%
\else
\language=\csname l@#1\endcsname
\fi
#2}}

\bibitem{antoulas2000survey}
A.~C. Antoulas, D.~C. Sorensen, and S.~Gugercin, ``A survey of model reduction
  methods for large-scale systems,'' Tech. Rep., 2000.

\bibitem{Baur14}
\BIBentryALTinterwordspacing
U.~Baur, P.~Benner, and L.~Feng, ``Model order reduction for linear and
  nonlinear systems: A system-theoretic perspective,'' \emph{Archives of
  Computational Methods in Engineering}, vol.~21, no.~4, pp. 331--358, 2014.
  [Online]. Available: \url{https://doi.org/10.1007/s11831-014-9111-2}
\BIBentrySTDinterwordspacing

\bibitem{Lassila14}
\BIBentryALTinterwordspacing
T.~Lassila, A.~Manzoni, A.~Quarteroni, and G.~Rozza, \emph{Model Order
  Reduction in Fluid Dynamics: Challenges and Perspectives}.\hskip 1em plus
  0.5em minus 0.4em\relax Cham: Springer International Publishing, 2014, pp.
  235--273. [Online]. Available:
  \url{https://doi.org/10.1007/978-3-319-02090-7_9}
\BIBentrySTDinterwordspacing

\bibitem{Hetmaniuk12}
\BIBentryALTinterwordspacing
U.~Hetmaniuk, R.~Tezaur, and C.~Farhat, ``Review and assessment of
  interpolatory model order reduction methods for frequency response structural
  dynamics and acoustics problems,'' \emph{International Journal for Numerical
  Methods in Engineering}, vol.~90, no.~13, pp. 1636--1662, 2012. [Online].
  Available: \url{https://onlinelibrary.wiley.com/doi/abs/10.1002/nme.4271}
\BIBentrySTDinterwordspacing

\bibitem{ionescu_astolfi_2011}
T.~C. Ionescu and A.~Astolfi, ``Moment matching for linear systems -- overview
  and new results*,'' \emph{IFAC Proceedings Volumes}, vol.~44, no.~1, pp.
  12\,739--12\,744, 2011.

\bibitem{antoulas1986scalar}
A.~Antoulas and B.~Anderson, ``On the scalar rational interpolation problem,''
  \emph{IMA Journal of Mathematical Control and Information}, vol.~3, no. 2-3,
  pp. 61--88, 1986.

\bibitem{Nakatsukasa_2018}
\BIBentryALTinterwordspacing
Y.~Nakatsukasa, O.~S{\`{e}}te, and L.~N. Trefethen, ``The {AAA} algorithm for
  rational approximation,'' \emph{{SIAM} Journal on Scientific Computing},
  vol.~40, no.~3, pp. A1494--A1522, jan 2018. [Online]. Available:
  \url{https://doi.org/10.1137%2F16m1106122}
\BIBentrySTDinterwordspacing

\bibitem{berrut_trefethen_2004}
J.-P. Berrut and L.~N. Trefethen, ``Barycentric lagrange interpolation,''
  \emph{SIAM Review}, vol.~46, no.~3, pp. 501--517, 2004.

\bibitem{gosea2021algorithms}
I.~V. Gosea and S.~G{\"u}ttel, ``Algorithms for the rational approximation of
  matrix-valued functions,'' 2021.

\bibitem{Rodriguez23}
\BIBentryALTinterwordspacing
A.~C. Rodriguez, L.~Balicki, and S.~Gugercin, ``The p-aaa algorithm for
  data-driven modeling of parametric dynamical systems,'' \emph{SIAM Journal on
  Scientific Computing}, vol.~45, no.~3, pp. A1332--A1358, 2023. [Online].
  Available: \url{https://doi.org/10.1137/20M1322698}
\BIBentrySTDinterwordspacing

\bibitem{gosea2021datadriven}
I.~V. Gosea and S.~Gugercin, ``Data-driven modeling of linear dynamical systems
  with quadratic output in the aaa framework,'' 2021.

\bibitem{yu2023leveraging}
A.~Yu and A.~Townsend, ``Leveraging the hankel norm approximation and block-aaa
  algorithms in reduced order modeling,'' 2023.

\bibitem{Bruinsma90}
\BIBentryALTinterwordspacing
N.~Bruinsma and M.~Steinbuch, ``A fast algorithm to compute the
  h{$\infty$}-norm of a transfer function matrix,'' \emph{Systems {\&} Control
  Letters}, vol.~14, no.~4, pp. 287--293, 1990. [Online]. Available:
  \url{https://www.sciencedirect.com/science/article/pii/016769119090049Z}
\BIBentrySTDinterwordspacing

\bibitem{Zhou95}
K.~G. Kemin~Zhou, John C.~Doyle, \emph{Robust and optimal control}.\hskip 1em
  plus 0.5em minus 0.4em\relax Prentice Hall, 1995.

\bibitem{iss_model}
\BIBentryALTinterwordspacing
Y.~Chahlaoui, ``Benchmark examples for model reduction.'' [Online]. Available:
  \url{http://slicot.org/20-site/126-benchmark-examples-for-model-reduction}
\BIBentrySTDinterwordspacing

\end{thebibliography}

\end{document}